\documentclass[a4paper, amsfonts, amssymb, amsmath, reprint, showkeys, nofootinbib, twoside]{revtex4-1}
\usepackage[english]{babel}
\usepackage[utf8]{inputenc}
\usepackage[colorinlistoftodos, color=green!40, prependcaption]{todonotes}
\usepackage{amsthm}
\usepackage{mathtools}
\usepackage{physics}
\usepackage{xcolor}
\usepackage{graphicx}
\usepackage[left=23mm,right=13mm,top=35mm,columnsep=15pt]{geometry} 
\usepackage{adjustbox}
\usepackage{placeins}
\usepackage[T1]{fontenc}
\usepackage{lipsum}
\usepackage{csquotes}
\usepackage[pdftex, pdftitle={Article}, pdfauthor={Author}]{hyperref} % For hyperlinks in the PDF
\begin{document}

\title{Cosmological Expansion with Global Phase Normalization by the Hubble Horizon}

\author{Maurice H.P.M. van Putten}
    \email{mvp@sejong.ac.kr}% Your name
\affiliation{INAF-Osservatorio Astronomico di Capodimonte, Salita Moiariello 16, I-80131 Napoli, Italy}
\affiliation{Department of Physics and Astronomy, Sejong University, 98 Gunja-Dong, Gwangjin-gu, Seoul 143-747, Republic of Korea}

\date{\today} % Leave empty to omit a date

\begin{abstract}
We argue that cosmological expansion is subject to a global phase normalization in the gravitational path integral, fixed by causal horizon boundary conditions rather than by local dynamics. In this formulation, the cosmological conformal factor is not a propagating degree of freedom but a global gauge variable fixed by the Hamiltonian constraint, rendering the conventional conformal-factor problem inapplicable. A de Sitter turning point $(q=-1)$ uniquely fixes the phase density $\Lambda = J$ as the leading order integrating factor when the horizon Clausius relation holds, where $J$ is the trace of the Schouten tensor of the cosmological background. 
We parameterize departures from equilibrium by a single variance parameter $\beta$ governing non-adiabatic background evolution over a Hubble time scale. 
The resulting Hubble expansion points to a phantom regime beyond $\Lambda$CDM without new degrees of freedom. It provides a natural origin for cosmological tensions expressed by cosmographic parameters, arising from global constraints that inhibit a stable de Sitter universe.
\end{abstract}


\maketitle

{\it Introduction---}The cosmological constant and the conformal factor problems in cosmological spacetime remain longstanding conceptual issues \citep{gib78,fel17,hal19}. 
The empirical success of $\Lambda$CDM in CMB analysis is tempered by persistent tensions with the Hubble constant of the late-time Universe \citep{div25}. 
In Euclidean approaches to quantum gravity, the conformal mode appears with a wrong-sign kinetic term, leading to the so-called conformal-factor problem. 
These two problems are usually treated as unrelated. 

In a path-integral formulation of quantum cosmology, the gravitational phase requires a running normalization determined by cosmological horizon as a global causal 
boundary to phase space within -- mirroring Mach’s principle \citep{bar95}. The conformal factor then appears as a global gauge variable fixed by the Hamiltonian constraint, 
rather than as a propagating degree of freedom, and cosmological expansion is subject to phase normalization in the variational principle extended to super-Hubble scale fluctuations.

With an appeal to Clausius' relation, the use of horizon {\it equilibrium} thermodynamics is frequently invoked to provide a global constraint on gravitational dynamics \citep{jac95,eas11}, 
and further developed in approaches emphasizing the balance between bulk and boundary degrees of freedom with possible cosmological implications \citep{pad10,ver11}. 
The present approach differs in identifying equilibrium thermodynamics with a turning point in the resulting Hubble expansion, rather than to generate local field equations.
Accordingly, the cosmological constant $\Lambda$ is identified with an equivalent phase density, inherently dynamic, rather than fundamental constant that would disappear  
upon differentiation of SO(3,1) formulations of the field equations \citep{van96}, 

A causal path integral on a horizon-bounded spacetime demands a global phase reference within the causal domain. 
In FLRW space–time, we thus encounter the propagator $e^{i(S-S_0)}$ with running gauge $S_0[{\cal H}]$ for the total phase of background zero–wavenumber modes 
associated with the Hubble horizon ${\cal H}$ rather than $e^{iS}$ given the static gauge zero of null infinity of Minkowski spacetime \citep{van20}.
The global phase normalization acts on the dimensionless densities such as $\Omega_\Lambda = \tfrac13 R_H^2\Lambda$, where $R_H=c/H$ denotes the Hubble radius given the Hubble parameter $H$ and velocity of light $c$, while $\Lambda$ serves as the integrating factor. 
This allows standard field-theoretic techniques to be applied without introducing new degrees of freedom and preserving the non-dynamical character of the conformal factor.
This distinction clarifies how $\Lambda$ may vary without representing a propagating degree of freedom, and why departures from the exact de Sitter evolution
may lead to an effective phantom behavior.

{\it Causality and Phase Normalization---}In a Lorentzian path-integral formulation of gravity, only phase differences are physically meaningful. 
On cosmological backgrounds, however, causality restricts correlations to lie within the cosmological horizon. 
As a result, the phase of the gravitational action must be defined relative to a global reference fixed by the causal domain itself. 
The cosmological horizon provides such a reference by delimiting the total causal phase space.

Local gravitational dynamics are defined relative to the global state of the Universe, defined by the variational principle for sub-horizon scale fluctuations
\citep{van24}.
The background spacetime hereto may evolve by conformal scaling, which carries a scaling dimension $\Delta = 2$ based on UV-IR-consistency of causal phase space with its horizon boundary \citep{van20,van26b}. Mirroring Mach, relativity is manifest on a Hubble scale as a nonlocal phase normalization without modification of the local Einstein field equations.
Once the gravitational path integral is restricted to a horizon-bounded causal domain, no local variational principle determines the overall phase of the action. 
Consistency with causality requires a global condition to act on the total causal phase space. 

Away from exact de Sitter symmetry, the global integrating factor becomes time dependent. Its leading behavior may be expressed in terms of cosmographic invariants by expanding the effective cosmological term as \citep{van26a}
\begin{eqnarray}
    \Lambda(a) \;=\; J(a) \;+3\; \beta\,(j-3) H^2/c^2\;+\; \cdots,
    \label{EQN_Lambda}
\end{eqnarray}
where $J(a)$ denotes the trace of the Schouten tensor and $j$ is the cosmographic jerk parameter.
The combination $j-3$ vanishes in the radiation-dominated era, and $\beta$ parametrizes delayed adjustment of the integrating factor to evolving causal phase space.
The coefficient $\beta$ may hereby parametrize departures from adiabatic evolution, providing a correction to 
phase normalization away from thermodynamic equilibrium.

{\it Conformal Factor and the Hamiltonian Constraint---}Mach’s principle elevated to a path-integral formulation within a finite causal domain 
predicts the existence of a spacetime-wide conformal scaling. 
A global phase reference fixed by the cosmological horizon must be consistently transported throughout the causal domain within. 
In spatially flat cosmology, this uniquely leads to a Friedmann line element $ds^2=a^2\eta_{ab}dx^adx^b$ in conformal time. 
Allowing the conformal scaling to evolve implies a Hubble parameter whereby cosmological expansion emerges as a consequence of phase normalization, 
rather than as an independent assumption. This evolution may include a {\it de Sitter turning point} $q=-1$, where $q$ is the deceleration parameter, 
when the horizon is momentarily stationary and the size of causal phase space is extremal \citep{van12}. 
While the de Sitter universe $\Omega_\Lambda= 1$ $(q\equiv -1)$ seals the fate of the Universe in $\Lambda$CDM, 
this need not be supported by the underlying causal phase space.

In homogeneous and isotropic cosmology, the metric is written as $ds^2 = -dt^2 + a^2(t)\,d\vec{x}^2$ with respect to cosmic time $t$. 
The scale factor $a(t)$ enters the theory exclusively through the Hamiltonian constraint, which fixes the closure density of the Universe. 
It does not correspond to a propagating degree of freedom and possesses no canonical conjugate momentum. 
Consequently, the conformal factor cannot be endowed with a conventional kinetic term, and interpretations of the conformal-factor 
problem as a dynamical instability are inapplicable in a cosmological setting. The conformal factor functions, instead, as a global gauge variable 
fixed by boundary conditions imposed at the cosmological horizon. Its conjugate quantity is not a canonical momentum but variable associated 
with horizon geometry, notably horizon surface gravity \citep{van15,van17,van24},
\begin{equation}
a_H = \tfrac{1}{2}(1-q)\,a_{\rm dS}, \qquad
a_{\rm dS} = \frac{cH}{2\pi},
\label{EQN_aH}
\end{equation}
where $a_{\rm dS}$ denotes the de Sitter limit $q=-1$ \citep{gib77}. 
The Hamiltonian constraint thus enforces global phase normalization, which satisfies thermodynamic consistency rather than through local stationarity of the action.

{\it Gibbs’ Principle at de Sitter turning points ---}
A de Sitter turning point $q=-1$ represents time symmetry rather than a dynamical fixed point.
In its neighborhood, the background is maximally symmetric and thermodynamically well defined, so that Gibbs' variational principle applies to virtual displacements at fixed total energy. 
Consistency of horizon thermodynamics then uniquely fixes the equilibrium phase density as the integrating factor that renders the horizon Clausius relation exact. 
Including analytic continuation away from the time-symmetric de Sitter turning point, this yields
\begin{equation}
\Lambda = J = \frac{1}{6}R,
\label{EQN_J}
\end{equation}
where $R$ is the Ricci scalar of the Friedmann background and $J$ is the trace of the Schouten tensor. 
The numerical coefficient is fixed at the de Sitter turning point, obeying thermodynamic equilibrium, not stability of a dynamical conformal mode.
In this formulation, $\Lambda$ is elevated from an integration constant to an integrating factor, fixing the global phase normalization of the causal path integral. 

{\it Departures from Equilibrium---}Away from turning points, the cosmological background may evolve non-adiabatically on a Hubble time scale.
In (\ref{EQN_Lambda}), we encode this by a single variance parameter $\beta$, which quantifies the statistical spread in the realization of the Machian phase reference as the causal phase space evolves. $\beta$ does not describe fluctuations of the conformal factor itself, but rather the delayed adjustment of the integrating factor to changes in horizon geometry.

Requiring that $\Lambda$ is negligible during the matter-dominated era selects an upper bound $\beta \lesssim \beta_{\rm crit} = 1/12$; 
values above $\beta_{\rm crit}$ would lead to $\Omega_\Lambda < 0$ during matter domination. 
For $\beta$ near this critical value, the integrating factor remains subdominant while the universe is matter dominated ($q \simeq 1/2$), but evolves non-adiabatically at later times. 
As a result, the effective dark-energy component eventually grows relative to the matter density, producing a crossing of the phantom divide ($w=-1$) at $z_c \simeq 1.85$ about cosmic noon, determined by a direct fit to the Planck CMB using CAMB \citep{van25b,van26a}.

{\it Comparison of $\Lambda$CDM versus $J$CDM---}
In $\Lambda$CDM, $\Lambda$ is assumed to be strictly constant, fixing the global phase density by hand. 
This choice is appropriate to early-time equilibrium under adiabatic evolution and yields an excellent description of early-universe observables. 
However, once the causal phase space evolves, this rigid phase assignment becomes inconsistent with Machian phase normalization.
In contrast, $J$CDM identifies $\Lambda$ with a phase density (\ref{EQN_J}) representing a running global phase normalization, 
analytically continued away from the de Sitter urning point.  As a result, $J$CDM naturally captures data from the Local Distance Ladder, 
where the relevant horizon scale is accurately represented, while exhibiting tension with early-universe observables when the background evolution may be non-adiabatic.

{\it Conclusions---}We argue that, mirroring Mach’s principle, the gravitational path integral for a Big Bang cosmology involves a global phase normalization by the cosmological horizon. 
The cosmological conformal factor is hereby fixed by the Hamiltonian constraint that acts as a global gauge variable, 
not a field with canonical momentum, thereby resolving the conformal-factor problem by construction.
This approach suggests a common origin to the cosmological constant problem, the conformal-factor problem, and late-time cosmological tensions.
While general relativity continues to describe sub-horizon geometric deformations, cosmological expansion of the cosmological background spacetime 
is subject to a dark-energy arise from a running global phase derived from the Hubble horizon as a global causal boundary to its phase space. 
It features a de Sitter turning point $(q=-1)$, where Gibbs’ principle applies \citep{bek73, bek81, haw75,eas11},
while rendering the de Sitter universe to be unstable, thus ruling out $\Lambda$CDM in the future with finite tensions in the present epoch. A full reconciliation of late- and early-time cosmology appears to involve small departures from equilibrium in $\beta J{\rm CDM}$ with $\beta$ close to $\tfrac{1}{12}$ \citep{van26a}.

{\it Acknowledgements---}This research is supported in part by National Research Foundation (NRF) grant No. RS-2024-00334550.

%\appendix*
%\input{sections/appendix1.tex}

\end{document}